\documentclass[lettersize,journal]{IEEEtran}
\usepackage{amsmath,amsfonts}
\usepackage{algorithmic}
\usepackage{algorithm}
\usepackage{array}
\usepackage[caption=false,font=normalsize,labelfont=sf,textfont=sf]{subfig}
\usepackage{textcomp}
\usepackage{stfloats}
\usepackage{url}
\usepackage{verbatim}
\usepackage{graphicx}
\usepackage{cite}
\usepackage{multirow}
\usepackage{amssymb}
\usepackage{booktabs}
\DeclareMathAlphabet{\mathbbold}{U}{bbold}{m}{n}
\begin{document}

\title{On the Role of Conversational Timing in \\ Synthetic Training Data for ASR}

\author{Máté Gedeon$^{*,\dagger}$, Péter Mihajlik$^{*}$ \\
\vspace{5pt}

    $^{*}$Dept. of Telecommunications and Artificial Intelligence, \\
    Budapest University of Technology and Economics, Hungary \\
    $^{\dagger}$Speechtex Ltd. \\
    \vspace{5pt}
    \texttt{gedeonm@edu.bme.hu, mihajlik@tmit.bme.hu}
}



\maketitle

\begin{abstract}
Synthetic multi-speaker conversations are widely used to train conversational automatic speech recognition (ASR) systems, but it remains unclear which timing properties make simulated data most useful. This paper studies conversational timing as a controllable training variable rather than merely as a corpus statistic to be reproduced. We parameterize pause and overlap timing distributions with an exponential-tilting family estimated from multiple conversational corpora, and then explore the resulting four-dimensional parameter space with Latin hypercube sampling and multi-objective Bayesian optimization. Each sampled timing configuration is used to generate simulated training conversations, train an ASR system, and evaluate concatenated-permutation word and character error rates (cpWER and cpCER) on a Hungarian dialogue corpus. The results show that downstream ASR behavior is explained more directly by induced timing statistics than by raw simulator coordinates or corpus proximity. In particular, higher overlap exposure is associated with lower cpWER, whereas longer and more variable gaps are associated with higher cpWER; cpCER follows the same trend, but with weaker statistical support. Bayesian optimization yields modest aggregate improvements, but its main value is analytical: it produces controlled timing interventions that reveal an overlap--gap trade-off in simulated conversational training data. These findings suggest that realistic simulation should be complemented by task-relevant diagnostics of overlap, gap, and timing-variability profiles.
\end{abstract}


\section{Introduction}
\IEEEPARstart{C}{onversational} speech simulation is widely used in modern speech technology when natural multi-speaker data are limited in size, diversity, or annotation quality \cite{AMI,CHiME-5,CHiME-6}. A common approach is to construct artificial conversations from single-speaker recordings, making it possible to generate large amounts of training data while controlling interaction properties such as pause duration, overlap frequency, and turn-taking behavior \cite{LibriCSS,Landini2022,Yamashita2022Naturalness}. These properties are more than descriptive statistics of a conversation: they affect segmentation ambiguity, acoustic interference, speaker attribution, and, ultimately, recognition and diarization performance \cite{SOT,Landini2022MultiSpeakerEEND,C-SASC}.

Most existing simulation pipelines are designed around a realism objective. They estimate timing statistics from conversational corpora and generate synthetic interactions that reproduce those observed distributions \cite{Landini2022,SASC,Yamashita2022Naturalness}. This is a natural and useful design principle, because unrealistic timing can produce training data that differ substantially from spontaneous conversations. However, realism alone does not answer a second question that is central to training: which timing properties make simulated data useful for a downstream model? A corpus-faithful distribution may be realistic, but it is not necessarily the most useful distribution for automatic speech recognition (ASR) or end-to-end neural diarization (EEND).

This distinction motivates the present work. Instead of treating simulation only as a way to imitate a target corpus, we treat conversational timing as a controllable object for systematic analysis. In particular, we ask how overlap rate, mean overlap duration, pause statistics, tail behavior, and the position of a distribution in a predefined parameter space relate to downstream ASR performance. The goal is not simply to identify a single best simulation setting, but to understand which properties of a simulated timing distribution are associated with lower or higher recognition error.

To support this analysis, we use a compact parameterization of conversational timing based on exponential tilting \cite{Esscher1932}. Starting from a base distribution estimated from multiple conversational corpora, the model generates a smooth family of related timing distributions through a low-dimensional parameter vector. This construction is useful for two reasons. First, it can represent corpus-like timing regimes. Second, it permits controlled deviations from those regimes, making it possible to probe regions of the timing space that are not tied to any single source corpus.

We then explore this parameter space using a practical hybrid strategy. Latin hypercube sampling (LHS) \cite{LHS} provides broad initial coverage, while Gaussian-process Bayesian optimization \cite{Jones1998EGO,Rasmussen2006GP,Shahriari2016BO} provides sample-efficient adaptive refinement under a limited training budget. In this paper, the optimization process is used not only as a search procedure, but also as an experimental design mechanism: every evaluated point becomes a controlled timing condition that can be analyzed in terms of both intrinsic statistics and ASR behavior.

We evaluate this framework on the Hungarian BEA-Dialogue benchmark \cite{bea_large}. The empirical study is organized around research questions rather than only around a best-score comparison. We analyze how concatenated-permutation word and character error rates (cpWER and cpCER) metrics, commonly used for multi-speaker conversational ASR evaluation \cite{CHiME-6,LibriCSS}, vary with overlap statistics, gap statistics, distributional distance from corpus references, and tilt-vector components. This makes the study a structured investigation of what makes a simulated conversational timing distribution useful for ASR training.

The main contributions of this work are summarized as follows:
\begin{itemize}
\item We introduce an exponential-tilting formulation that yields a compact, interpretable, and controllable family of conversational timing distributions anchored in multi-corpus conversational data.

\item We frame conversational simulation as a structured analysis problem, asking which intrinsic timing properties are associated with downstream ASR behavior rather than relying only on aggregate best-configuration comparisons.

\item We define a measurement protocol for simulated conversations, including overlap rate, mean overlap, gap statistics, tail mass, boundary density, distributional distance to corpus references, and tilt-space position.

\item We combine LHS and Bayesian optimization to sample informative regions of the timing space and use the resulting evaluations to study relationships between timing parameters, intrinsic timing statistics, and cpWER/cpCER.

\item We provide an analysis-oriented case study on BEA-Dialogue that connects simulation design choices to ASR error patterns.
\end{itemize}

The remainder of the paper is organized as follows. Section~II reviews data augmentation and conversational simulation methods. Section~III describes the timing parameterization and optimization procedure. Section~IV presents the experimental protocol and research questions. Section~V reports the analysis of timing statistics, parameter-space structure, and ASR errors. Sections~VI and VII discuss the implications and conclude the paper.

\section{Related Work}
Data augmentation has long been used to improve ASR robustness, with methods such as vocal-tract-length perturbation, speed perturbation, additive noise, reverberation, and time-frequency masking \cite{Jaitly2013VTLP,ko15_interspeech,Cui_augmentation,Park2019}. Although highly effective, these techniques primarily target channel and environmental variability. They do not model conversational interaction structure, and therefore offer limited control over temporal phenomena that are central to multi-speaker speech, such as pause patterns, overlap incidence, and turn-exchange dynamics.

To address this gap, prior work on synthetic conversation generation has constructed multi-speaker sessions from single-speaker recordings. Early pipelines relied on random concatenation with heuristic pauses \cite{Fujita2019}, which increased the amount of training data but often produced unrealistic interaction timing relative to continuous conversational recordings, where overlap and non-overlap regions coexist \cite{LibriCSS}. More recent approaches introduced statistically grounded simulation: pause and overlap distributions are estimated from real conversations and used to generate more natural turn-taking behavior \cite{Yamashita2022Naturalness,Landini2022}. Speaker-aware frameworks further improve realism by incorporating speaker-dependent timing characteristics and Markovian turn models \cite{SASC,C-SASC}. These advances have demonstrated clear downstream benefits for diarization and conversational ASR \cite{Landini2022MultiSpeakerEEND,C-SASC}.

Despite this progress, most existing methods are designed primarily for realism, typically measured as proximity to empirical corpus statistics. This emphasis leaves a central question insufficiently explored: whether corpus-matching timing distributions are also optimal for model training. From a learning perspective, the objective is not to replicate data statistics per se, but to produce training conditions that minimize recognition error on the target task. In challenging conversational settings, these two objectives can diverge.

The broader machine-learning literature provides a related perspective. Distribution shaping through reweighting, resampling, or cost-sensitive design is widely used to improve generalization under dataset shift, class imbalance, and domain mismatch \cite{Shimodara2000,He2009,Japkowicz2002,Sugiyama2007CovariateShift}. These methods share the principle of modifying the effective training distribution so that learning focuses on informative regions of the sample space. However, they are typically applied at the sample or label level; they do not directly provide a structured mechanism for manipulating conversational timing statistics in synthetic speech generation.

Our work builds on the realism-oriented simulation literature while addressing this missing link. Instead of treating conversational timing as a fixed quantity inferred from a single corpus, we model it as a controllable distribution family and optimize its parameters against downstream ASR performance. This framing motivates the methodology in the next section: a compact exponential-tilting parameterization for timing distributions, coupled with sample-efficient black-box optimization over the resulting parameter space.

\section{Research Questions}

We organize the experimental study around the following research questions:

\begin{enumerate}
\item[\textbf{RQ1}] Which intrinsic timing statistics (overlap, tail mass, etc.) of simulated conversations are most associated with cpWER and cpCER?
\item[\textbf{RQ2}] Do high-performing settings lie near corpus-derived timing distributions, near the mixture base distribution, or in extrapolated regions of the tilt space?
\item[\textbf{RQ3}] Which components of the tilt vector $\theta$ are most informative for downstream ASR performance?
\item[\textbf{RQ4}] To what extent is the used parameter space redundant?
\item[\textbf{RQ5}] How efficiently does the optimization framework improve ASR performance?
\end{enumerate}

Together, these questions emphasize explanation, with the objective of identifying what makes a timing distribution useful for training, not merely reporting the lowest observed error rate.

\section{Methodology}
We seek a controllable family of conversational timing distributions and an optimization procedure that identifies simulation settings yielding strong downstream ASR performance. Let $\delta \in \mathbb{R}$ denote the inter-utterance timing variable: $\delta < 0$ corresponds to overlap, and $\delta \ge 0$ corresponds to a silence gap. For a simulation configuration parameterized by $\theta$, training and evaluation produce a stochastic performance vector
\begin{equation}
\mathbf{f}(\theta)=\big(f_1(\theta), f_2(\theta)\big)=\big(\mathrm{cpWER}(\theta), \mathrm{cpCER}(\theta)\big),
\end{equation}
and the objective is to minimize both.

At a high level, the methodology has two stages. First, we build an interpretable parametric model of conversational timing that can move smoothly between different timing regimes. Second, we search this parameter space with a sample-efficient optimizer, where each parameter setting is scored by end-to-end ASR performance.

\subsection{Parameterizing Conversational Timing Distributions}
\subsubsection{Timing variable and decomposition}
Following the speaker-aware simulated conversation (SASC) framework \cite{SASC}, the timing process is decomposed into a mean component and a residual component. Let $X_n$ denote the speaker of utterance $n$. The inter-utterance timing is modeled as
\begin{equation}
\delta_n=
\begin{cases}
\mu^{=}_{s,n}+v_n, & X_n=X_{n-1},\; v_n\sim V_{=},\\[2pt]
\mu^{\neq}_{s,n}+v_n, & X_n\neq X_{n-1},\; v_n\sim V_{\neq},
\end{cases}
\end{equation}
where $\mu^{=}_{s,n}$ and $\mu^{\neq}_{s,n}$ are speaker-dependent mean terms, and $V_{=},V_{\neq}$ capture local residual variability.

Intuitively, the mean term determines the coarse conversational rhythm (the average amount of overlap and the average pause duration), while the residual term captures fine-grained randomness around that rhythm.

In this work, we modify only the distribution of the mean timing term and keep the residual model $(V_{=},V_{\neq})$ identical to that in the original SASC implementation. This isolates the effect of global timing statistics (overlap probability and pause-length profile), while preserving the speaker-aware capabilities of the framework.

\subsubsection{Corpus density estimation}
Assume a set of $K$ conversational corpora. For corpus $k$, let $p_k(\delta)$ be the empirical density of mean timing values. We estimate $p_k$ using kernel density estimation (KDE) \cite{Rosenblatt1956,Parzen1962}, with bandwidth selected by Silverman's rule \cite{silverman1986}. Each density is represented on a shared support grid to enable stable numerical optimization and comparison across corpora. This step converts each corpus into a comparable timing profile that can be manipulated mathematically.

\subsubsection{Exponential-tilting family}
We define a base density $p_0$ and use exponential tilting to generate a parametric family \cite{Esscher1932}:
\begin{equation}
p_{\theta}(\delta)=\frac{p_0(\delta)\exp\!\big(\theta^\top T(\delta)\big)}{Z(\theta)},
\label{eq:tilted_density}
\end{equation}
with normalizing constant
\begin{equation}
Z(\theta)=\int p_0(u)\exp\!\big(\theta^\top T(u)\big)\,du,
\label{eq:normalizer}
\end{equation}
where $\theta\in\mathbb{R}^4$ and
\begin{equation}
T(\delta)=
\begin{bmatrix}
\mathbbold{1}(\delta<0)\\
\delta\\
\max(0,\delta)\\
\min(0,\delta)
\end{bmatrix}.
\label{eq:features}
\end{equation}

The first statistic controls the relative mass assigned to overlapping transitions. The remaining statistics provide complementary control over the global displacement of the delay distribution and over the positive-gap and negative-overlap sides. Thus, the tilted family in \eqref{eq:tilted_density}--\eqref{eq:features} defines a smooth, low-dimensional set of distributions that remains absolutely continuous with respect to $p_0$.

The statistic vector in \eqref{eq:features} is intentionally overcomplete because
\begin{equation}
\delta = \max(0,\delta) + \min(0,\delta),
\end{equation}
so the second component is linearly dependent on the third and fourth components. Consequently, the induced density is identifiable only through the combinations
\begin{equation}
\phi_1=\theta_1,\qquad
\phi_2=\theta_2+\theta_3,\qquad
\phi_3=\theta_2+\theta_4.
\end{equation}
Equivalently, $\phi_2$ controls the effective slope on the positive-gap side, whereas $\phi_3$ controls the effective slope on the negative-overlap side. We nevertheless retain the four-coordinate representation as an experimental-design parameterization because its coordinates correspond to conceptually distinct simulator interventions: overlap mass, global delay shift, gap-side modulation, and overlap-side modulation. In analyses where coordinate-level interpretation is useful, we report results in this overcomplete representation. When distributional identifiability or geometric comparisons are required, we instead use the compressed three-dimensional representation $\phi=(\phi_1,\phi_2,\phi_3)$ or distribution-level quantities such as intrinsic timing statistics and KL divergence.

This distinction is important because the bounded search region used later is defined in the four-coordinate design space. Although the density itself is constant along one algebraic null direction, the four-dimensional box induces a particular feasible region and sampling geometry over the identifiable three-dimensional family. Therefore, the overcomplete representation should be understood as defining a bounded intervention space for simulation and optimization rather than as a minimal identifiable exponential-family parameterization.

This parameterization is useful because small changes in the design coordinates induce controlled and interpretable changes in the generated timing distributions. It therefore provides a substantially more stable and structured basis for downstream optimization than an unconstrained search over arbitrary densities.

\subsubsection{Base-density construction}
Rather than choosing a single reference corpus, we construct
\begin{equation}
p_0(\delta)=\sum_{k=1}^{K} w_k p_k(\delta),
\quad w_k\ge 0,\; \sum_{k=1}^{K} w_k=1.
\label{eq:base_mixture}
\end{equation}
Weights are chosen by minimizing the average projection error of corpus densities onto the tilted family using Kullback--Leibler (KL) divergence \cite{KullbackLeibler1951}:
\begin{equation}
\min_{\mathbf{w}\in\Delta^{K-1}}\frac{1}{K}\sum_{k=1}^{K}
D_{\mathrm{KL}}\!\left(p_k\,\|\,p_{\theta_k^{\star}(\mathbf{w})}\right),
\label{eq:weight_opt}
\end{equation}
where the corpus-specific projection is
\begin{equation}
\theta_k^{\star}(\mathbf{w})=\arg\min_{\theta}
D_{\mathrm{KL}}\!\left(p_k\,\|\,p_{\theta}\right),
\label{eq:theta_projection}
\end{equation}
and $\Delta^{K-1}$ denotes the simplex.

Conceptually, $p_0$ acts as a shared anchor across corpora. The optimization in \eqref{eq:weight_opt} selects mixture weights so that each corpus can be well represented after tilting, instead of forcing all corpora to match one arbitrary reference domain.

\subsubsection{Corpus embedding and simulation map}
For a fixed $p_0$, each corpus is embedded into parameter space via \eqref{eq:theta_projection}, yielding $\{\theta_k^{\star}\}_{k=1}^{K}$. For any candidate $\theta$, we instantiate the simulator by sampling mean timing values from $p_{\theta}$ and then adding residual terms from $V_{=},V_{\neq}$ as in the SASC pipeline.

Thus, each point in parameter space corresponds to a concrete simulation policy, and moving in this space has a direct and measurable effect on generated conversation timing.

\subsection{Parameter-Space Exploration and Optimization}
\subsubsection{Search domain and initialization}
Let
\begin{equation}
\theta_i^{\min}=\min_k (\theta_k^{\star})_i,\qquad
\theta_i^{\max}=\max_k (\theta_k^{\star})_i.
\end{equation}
We then define a conservative expanded box
\begin{equation}
\Theta=\prod_{i=1}^{4}[\tilde\theta_i^{\min},\tilde\theta_i^{\max}],
\end{equation}
with
\begin{equation}
\tilde\theta_i^{b}=
\begin{cases}
\theta_i^{b} + \big(2\mathbbold{1}[b=\max]-1\big), & |\theta_i^{b}|<1,\\
1.5\,\theta_i^{b}, & |\theta_i^{b}|\ge 1,
\end{cases}
\quad b\in\{\min,\max\}.
\label{eq:expanded_bounds}
\end{equation}
We draw an initial design $\{\theta^{(i)}\}_{i=1}^{N_0}\sim\mathrm{LHS}(\Theta)$ \cite{LHS}, evaluate each point through full simulation $\rightarrow$ ASR training $\rightarrow$ validation, and obtain the initial observations
\begin{equation}
\mathcal{D}_{N_0}=\{(\theta^{(i)},\mathbf{f}(\theta^{(i)}))\}_{i=1}^{N_0}.
\end{equation}

The role of LHS is to provide broad initial coverage before model-based search begins; this reduces the risk of early bias toward a narrow region of the parameter space.

\subsubsection{Multi-objective Bayesian optimization}
After initialization, we run multi-objective Bayesian optimization to improve sample efficiency \cite{Jones1998EGO,Shahriari2016BO}.

Because each evaluation requires full ASR training, direct exhaustive search is infeasible. Bayesian optimization addresses by learning a surrogate of the objective landscape and querying only the most informative candidate points \cite{Shahriari2016BO}.

\paragraph{Surrogates.}
Each objective is modeled by an independent Gaussian process \cite{Rasmussen2006GP}:
\begin{equation}
f_j(\theta)\sim\mathcal{GP}(m_j(\theta),k_j(\theta,\theta')),
\quad j\in\{1,2\},
\end{equation}
with Mat\'ern covariance \cite{Matern1960,Stein1999}. Given observations $\mathcal{D}_t$, the posterior prediction at $\theta$ is
\begin{equation}
f_j(\theta)\mid\mathcal{D}_t\sim\mathcal{N}(\mu_j(\theta),\sigma_j^2(\theta)).
\end{equation}

Here, $\mu_j(\theta)$ summarizes predicted performance, and $\sigma_j(\theta)$ quantifies uncertainty; both are needed to trade off exploitation and exploration.

\paragraph{Pareto set and acquisition.}
Using the standard Pareto dominance relation for multi-objective optimization \cite{Miettinen1999}, $\theta_a$ dominates $\theta_b$ under minimization if
\begin{equation}
f_j(\theta_a)\le f_j(\theta_b)\;\forall j,\qquad
\exists j: f_j(\theta_a)<f_j(\theta_b).
\end{equation}
Let $\mathcal{P}_t$ be the non-dominated set in $\mathcal{D}_t$. We select the next candidate by maximizing expected hypervolume improvement (EHVI), which is based on the hypervolume indicator for Pareto-front quality \cite{ZitzlerThiele1999,Emmerich2006EHVI}:
\begin{equation}
\theta_{t+1}=\arg\max_{\theta\in\Theta}\alpha_{\mathrm{EHVI}}(\theta\mid\mathcal{D}_t).
\end{equation}
This criterion balances exploitation of low predicted error and exploration of uncertain regions while respecting the cpWER--cpCER trade-off.

Operationally, we avoid collapsing cpWER and cpCER into a single scalar; instead, we preserve the trade-off structure and search for non-dominated solutions.

\paragraph{Update.}
After evaluating $\theta_{t+1}$, we append $(\theta_{t+1},\mathbf{f}(\theta_{t+1}))$ to $\mathcal{D}_t$, refit the surrogates, and repeat this process until the evaluation budget is exhausted. The final output is the estimated Pareto set and the corresponding simulation configurations.

In summary, the optimization loop alternates between (i) fitting a probabilistic model of ASR performance over timing parameters and (ii) selecting the next experiment that is expected to improve the current Pareto front most efficiently.

\begin{algorithm}[t]
\caption{Optimization of Conversational Timing Parameters}
\label{alg:mobo_simulation}
\begin{algorithmic}[1]
\REQUIRE Corpora $\{C_k\}_{k=1}^K$, evaluation budget $B$, initial size $N_0$
\STATE Estimate corpus timing densities $\{p_k\}_{k=1}^K$ (KDE)
\STATE Solve \eqref{eq:weight_opt} to obtain base density $p_0$ in \eqref{eq:base_mixture}
\STATE For each $k$, compute corpus embedding $\theta_k^{\star}$ via \eqref{eq:theta_projection}
\STATE Construct expanded search box $\Theta$ using \eqref{eq:expanded_bounds}
\STATE Draw $\{\theta^{(i)}\}_{i=1}^{N_0}\sim\mathrm{LHS}(\Theta)$ and evaluate to form $\mathcal{D}_{N_0}$
\FOR{$t=N_0,\dots,B-1$}
  \STATE Fit GP surrogates for cpWER and cpCER using $\mathcal{D}_t$
  \STATE Select $\theta_{t+1}=\arg\max_{\theta\in\Theta}\alpha_{\mathrm{EHVI}}(\theta\mid\mathcal{D}_t)$
  \STATE Evaluate $\mathbf{f}(\theta_{t+1})$ via simulation, ASR training, and validation
  \STATE Update $\mathcal{D}_{t+1}=\mathcal{D}_t\cup\{(\theta_{t+1},\mathbf{f}(\theta_{t+1}))\}$
\ENDFOR
\STATE \textbf{return} non-dominated set extracted from $\mathcal{D}_B$
\end{algorithmic}
\end{algorithm}

\section{Experiments}

The experiments use the evaluated simulation settings to probe the conversational timing space in a structured way. Each setting is treated as a controlled intervention on the timing distribution, and the analysis asks which timing properties are associated with better or worse ASR behavior. This framing emphasizes interpretation: the purpose is to understand what makes a timing distribution useful for training, while still allowing a single configuration to be selected based on its aggregate error rate.

\subsection{Data and Evaluation Protocol}

The base timing distribution is constructed from three conversational corpora: BEA-Dialogue (Hungarian) \cite{bea_large}, CallHome (English) \cite{CallHome}, and GRASS (Austrian German) \cite{grass}. For each corpus, we estimate pause and overlap timing densities using the preprocessing and KDE pipeline described in \cite{C-SASC}. These corpus-level densities are then mapped to the shared timing support used by the exponential-tilting model.

ASR evaluation is performed on the BEA-Dialogue development and evaluation splits, while timing statistics used for model construction are extracted only from the BEA-Dialogue training partition. No development or evaluation timing annotations are used when fitting the timing family or choosing simulation parameters. This separation is important because the analysis concerns general timing properties, not direct fitting to the evaluation set.

For each evaluated parameter setting $\theta$, we generate simulated conversational training data, train the ASR system using the same recipe, and record concatenated-permutation word error rate (cpWER) and concatenated-permutation character error rate (cpCER) \cite{CHiME-5}. Training and fine-tuning were performed with the NVIDIA NeMo 2.6.2 toolkit \cite{nemo}. To maintain consistency with our previous studies, all experiments used the same experimental protocol, with the English FastConformer Large CTC model \cite{fastconformer}\footnote{\url{https://huggingface.co/nvidia/stt_en_fastconformer_ctc_large}} as the initialization checkpoint. Each model was trained on a single \textit{NVIDIA RTX 5000 Ada Generation} GPU (32 GB VRAM) with a batch size of 16, an initial learning rate of $5 \times 10^{-4}$, and a cosine annealing learning-rate schedule.

\subsection{Timing Family and Search Domain}

Following the methodology described above, the pooled base density is represented as a weighted mixture of corpus-specific densities, and each corpus is embedded into the four-dimensional tilt space. The fitted mixture has an average KL divergence of $0.02392$ across the three corpus projections. Table~\ref{tab:corpus_params} summarizes the corpus-level results.

\begin{table}[h]
\centering
\caption{Corpus-derived tilt parameters used to define the search region.}
\label{tab:corpus_params}
\resizebox{0.95\columnwidth}{!}{
\begin{tabular}{lccc}
\toprule
\textbf{Corpus} & \textbf{Weight} & $\boldsymbol{\theta}$ & \textbf{KL} \\
\midrule
GRASS & $0.208$ & $[-0.001,\; 2.794,\; -4.431,\; 7.225]$ & $0.034$ \\
CallHome & $0.718$ & $[-0.019,\; -0.244,\; 0.610,\; -0.854]$ & $0.014$ \\
BEA-Dialogue & $0.074$ & $[-0.367,\; -0.924,\; -1.032,\; 0.108]$ & $0.024$ \\
\bottomrule
\end{tabular}}
\end{table}

The corpus embeddings also illustrate why the overcomplete four-coordinate design is useful in practice. Although the density is identifiable only through the combinations $\phi_1=\theta_1$, $\phi_2=\theta_2+\theta_3$, and $\phi_3=\theta_2+\theta_4$, the fitted corpus parameters show that the redundant coordinates separate qualitatively different timing effects. For example, GRASS has a large positive global-delay coefficient ($\theta_2=2.794$), but this is partly counteracted on the positive-gap side by a strong negative gap-side term ($\theta_3=-4.431$), yielding an effective positive-side slope of $\phi_2=-1.637$. At the same time, this global term is reinforced on the negative-overlap side by a large positive overlap-side term ($\theta_4=7.225$), yielding $\phi_3=10.019$. In contrast, CallHome and BEA-Dialogue occupy more moderate regions, with effective coordinates $\phi=(-0.019, 0.366, -1.098)$ and $\phi=(-0.367, -1.956, -0.816)$, respectively. Thus, the four-dimensional representation exposes whether a corpus-like timing profile is obtained through a global shift, a gap-side correction, an overlap-side correction, or a combination of these effects. This is useful for defining the experimental search region: the raw coordinates span substantially different ranges ($\theta_2$ from $-0.924$ to $2.794$, $\theta_3$ from $-4.431$ to $0.610$, and $\theta_4$ from $-0.854$ to $7.225$), while the low projection KL values ($0.014$--$0.034$) indicate that the overcomplete parameterization still provides accurate corpus-level representations. We therefore use the four-dimensional coordinates to define and interpret simulator interventions, while relying on the compressed coordinates and distribution-level statistics when non-redundant geometric comparisons are required.

The learned embeddings define the empirical range of corpus-like timing. We expand that range using \eqref{eq:expanded_bounds}, yielding the search domain
\begin{equation}
\begin{aligned}
\theta_1 &\in [-1.37,\,1.00], \\
\theta_2 &\in [-1.92,\,4.19], \\
\theta_3 &\in [-6.65,\,1.61], \\
\theta_4 &\in [-1.85,\,10.84].
\end{aligned}
\end{equation}
This design allows us to compare corpus-like settings with controlled extrapolations outside the observed corpus embeddings.

\subsection{Intrinsic Timing Measurements}

For every generated training set, we compute intrinsic timing measures before ASR training. These measures describe the simulated conversations independently of recognition performance. The main quantities used in the analysis are the following:

\begin{itemize}
  \item Overlap rate ($\Pr(\delta<0)$), i.e., the amount of simultaneous speech exposure.
  \item Mean overlap ($\mathbb{E}[-\delta\mid\delta<0]$), i.e., the average duration of overlapped regions.
  \item Mean gap ($\mathbb{E}[\delta\mid\delta\ge 0]$), i.e., the average silence between adjacent utterances.
  \item Overlap tail mass ($\Pr(\delta \le q_{0.20})$), i.e., the proportion of delays corresponding to the largest overlaps (the lowest 20\% of delays across all experiments), where $q_{0.20}$ is the 20th percentile.
  \item Gap tail mass ($\Pr(\delta \ge q_{0.80})$), i.e., the proportion of delays corresponding to the largest gaps (the highest 20\% of delays across all experiments), where $q_{0.80}$ is the 80th percentile.
  \item Delay mean ($\mathbb{E}[\delta]$), i.e., the mean of all inter-utterance delays.
  \item Delay standard deviation ($\sigma[\delta]$), i.e., the standard deviation of all inter-utterance delays.
\end{itemize}
$q_{0.20}$ and $q_{0.80}$ are fixed global thresholds computed from all generated inter-utterance delays; in our data, $q_{0.20} < 0 < q_{0.80}$.

In addition, for each experiment, we include features that quantify the Euclidean distance between its $\phi$ vector and the reference $\phi$ vectors corresponding to the three source corpora and the mixture-based reference. We also include four features measuring the Kullback--Leibler divergence between the experiment-specific distribution and each of these reference distributions.

\begin{figure*}[t]
    \centering
    \includegraphics[width=0.6\linewidth]{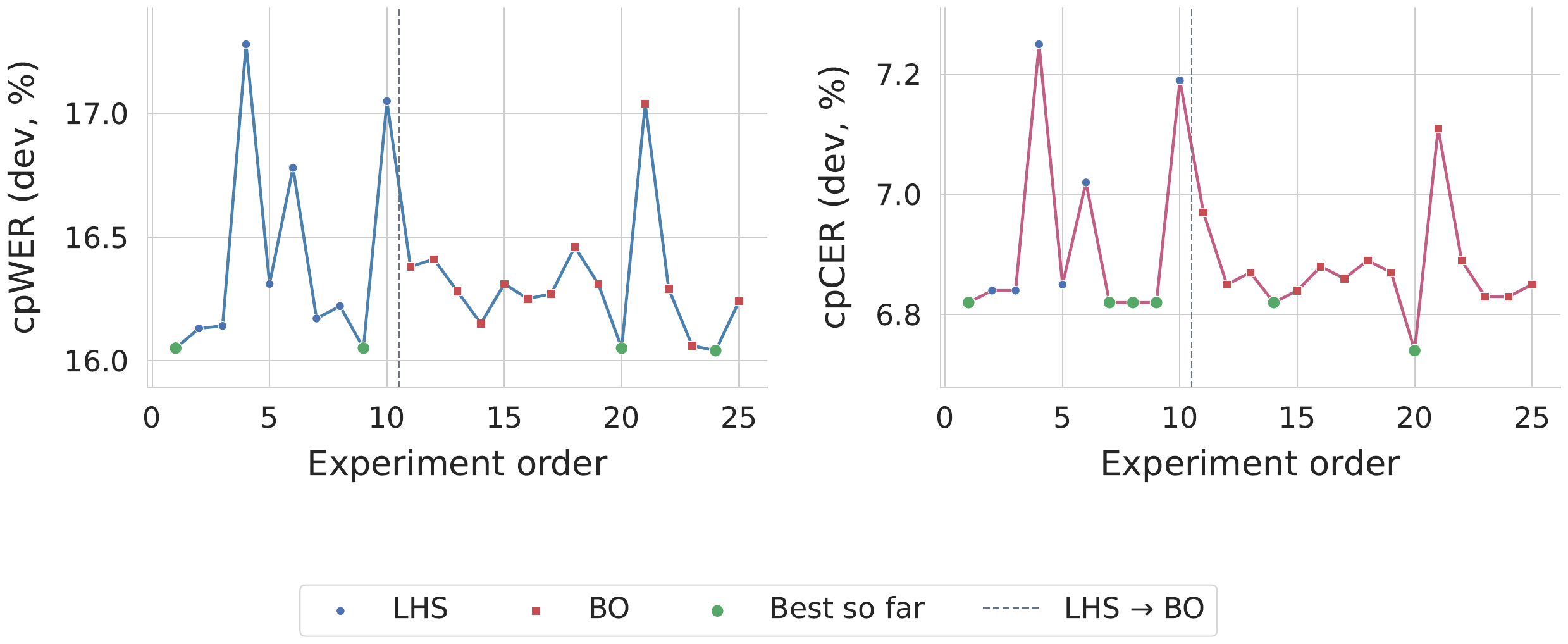}
    \caption{Overview of cpWER and cpCER across the LHS and Bayesian-optimization phases.}
    \label{fig:phase_overview}
\end{figure*}

\subsection{Analysis Procedure}
\label{sec:procedure}

The analysis proceeds in five steps, separating optimization behavior, distribution-level timing effects, and parameter-space geometry. This separation is important because the simulator searches in the overcomplete four-coordinate design space $\theta$, whereas the induced timing density is identifiable through the compressed coordinates $\phi=(\phi_1,\phi_2,\phi_3)$ and through observable timing statistics. We therefore keep these views as distinct as possible.

First, we present diagnostic plots that illustrate the optimization process, showing the evolution of cpWER and cpCER throughout the LHS and Bayesian-optimization phases. These results are compared with baseline and previously reported configurations to assess the effectiveness of the optimization procedure (\textit{RQ5}).

Next, we compare performance on the development and evaluation sets to investigate the generalization properties of the optimized configurations. Pearson and Spearman correlation analyses \cite{Pearson1895,Spearman1904} between the two splits are complemented by distribution visualizations for selected experiments, highlighting both high-performing and low-performing parameter settings.

To address \textit{RQ1} and \textit{RQ3}, we analyze the relationships between intrinsic timing measures, $\theta$ dimensions (and $\phi$ dimensions), and the target metrics. In addition to correlation coefficients with Benjamini--Hochberg correction for multiple testing \cite{BenjaminiHochberg1995}, we fit random-forest regression models \cite{Breiman2001}. These models are summarized with cross-validated $R^2$ and RMSE \cite{Stone1974}, along with permutation importance, to capture potentially nonlinear dependencies that may not be revealed by standard correlation analyses.

We then investigate \textit{RQ2} using scatter-plot visualizations, examining the relationships between key variables and performance metrics.

Finally, to address \textit{RQ4}, we analyze the structure of the $\theta$ and $\phi$ spaces using principal component analysis (PCA) \cite{Jolliffe2002PCA} to quantify their redundancy. We further apply clustering in the reduced two-dimensional space to explore whether distinct regions correspond to different performance characteristics and optimization behaviors.

\section{Results}
\label{sec:results}

This section evaluates the 25 generated timing configurations as a structured sample of the simulation-design space. The main purpose is not to argue from a single best configuration, but to identify which properties of the generated distributions are consistently associated with better or worse ASR performance. To keep the presentation focused, most tables and figures include only results from the evaluation set, while the development-set results are used as a consistency check and are discussed in the text where they affect the interpretation. A comprehensive set of numerical results with additional figures is made available in an online repository\footnote{\url{https://github.com/gedeonmate/conversation-timing}}. In the analyses below, we rely on several types of statistical evidence, as discussed in Section~\ref{sec:procedure}. Each analysis answers a different question, so the numbers should be interpreted together rather than as independent claims.

\subsection{Optimization Trace and Reliability of the Evaluation (\textit{RQ5})}
\label{sec:results_phase_consistency}

The experiments consist of 10 Latin hypercube sampling (LHS) configurations followed by 15 Bayesian optimization (BO) configurations. The LHS phase is intended to cover the search region broadly, whereas the BO phase uses previous ASR evaluations to choose new configurations. Fig.~\ref{fig:phase_overview} displays the evolution of cpWER and cpCER during the search, measured on the development set because this split guided the search. As expected, the LHS-phase results vary irregularly because they come from different regions of the $\theta$ space. During the BO phase, the trajectory shows a stronger exploitation trend, with occasional exploration steps (e.g., step 21).

On the evaluation set, the median cpWER decreases from 17.72 in the LHS phase to 17.61 in the BO phase (Mann--Whitney $p=0.037$ \cite{MannWhitney1947}), and the median cpCER decreases from 8.24 to 8.15 ($p<0.001$). Because BO points are adaptively selected, these p-values should be interpreted as diagnostics rather than as evidence from independent random groups. The descriptive comparison suggests that the adaptive phase moved the search into a somewhat more favorable region of the timing space. This caution is especially important because the development set gives a different picture: its medians do not improve under BO (cpWER: 16.20 for LHS and 16.28 for BO, $p=0.846$; cpCER: 6.84 for LHS and 6.86 for BO, $p=0.596$). This mismatch is informative: it warns against drawing conclusions from a single split or from a single best point. The rest of the section therefore emphasizes repeated patterns across metrics and analyses.

\begin{figure}[h]
    \centering
    \includegraphics[width=0.9\linewidth]{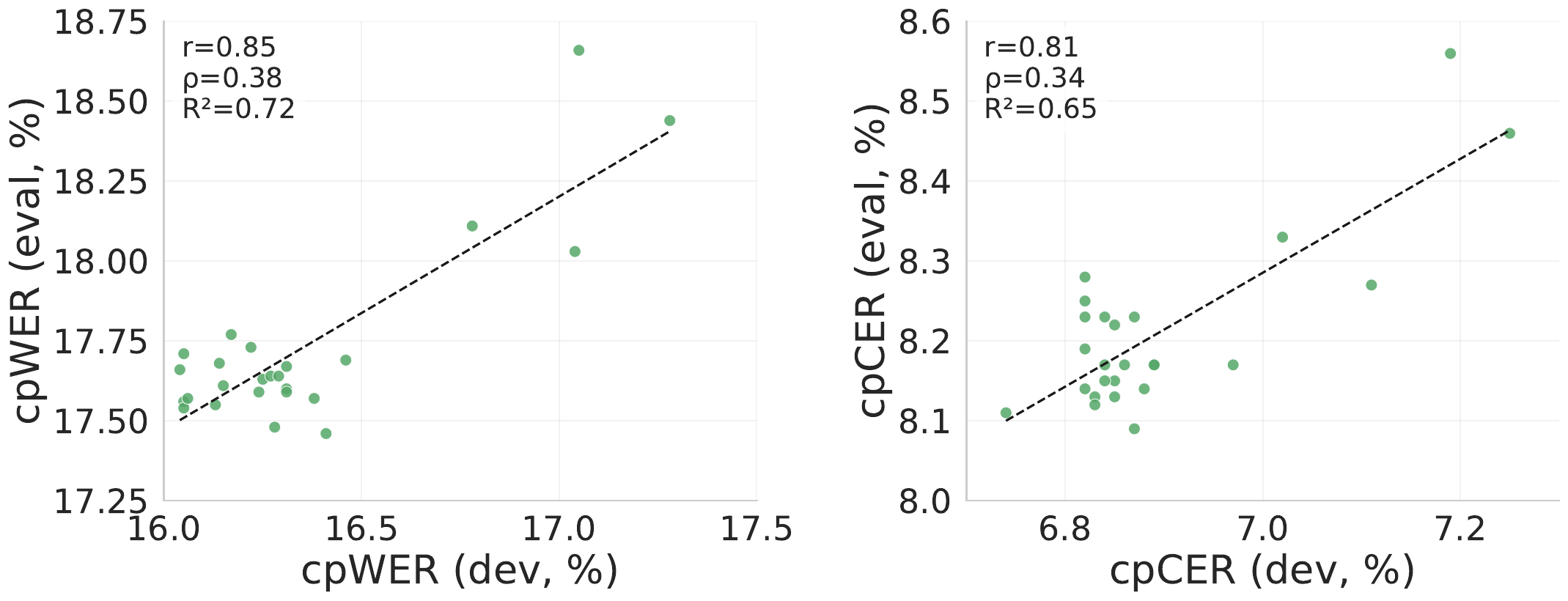}
    \caption{Development--evaluation consistency across configurations.}
    \label{fig:dev_eval_consistency}
\end{figure}

We next check whether development-set behavior is informative about evaluation-set behavior, as visualized in Fig.~\ref{fig:dev_eval_consistency}. Pearson correlation measures linear association between the error values themselves, while Spearman rank correlation measures whether the ordering of configurations is preserved \cite{Pearson1895,Spearman1904}. The development and evaluation errors have strong linear agreement (cpWER: Pearson $r=0.849$, $R^2=0.720$; cpCER: Pearson $r=0.805$, $R^2=0.649$), but weaker rank agreement (Spearman $\rho=0.375$ for cpWER and $\rho=0.343$ for cpCER). Thus, development results capture broad trends, but they are less reliable for ranking individual configurations. We therefore use the evaluation set for the main statistical interpretation.

\begin{figure*}[ht]
    \centering
    \begin{minipage}{0.4\textwidth}
        \centering
        \includegraphics[width=\linewidth]{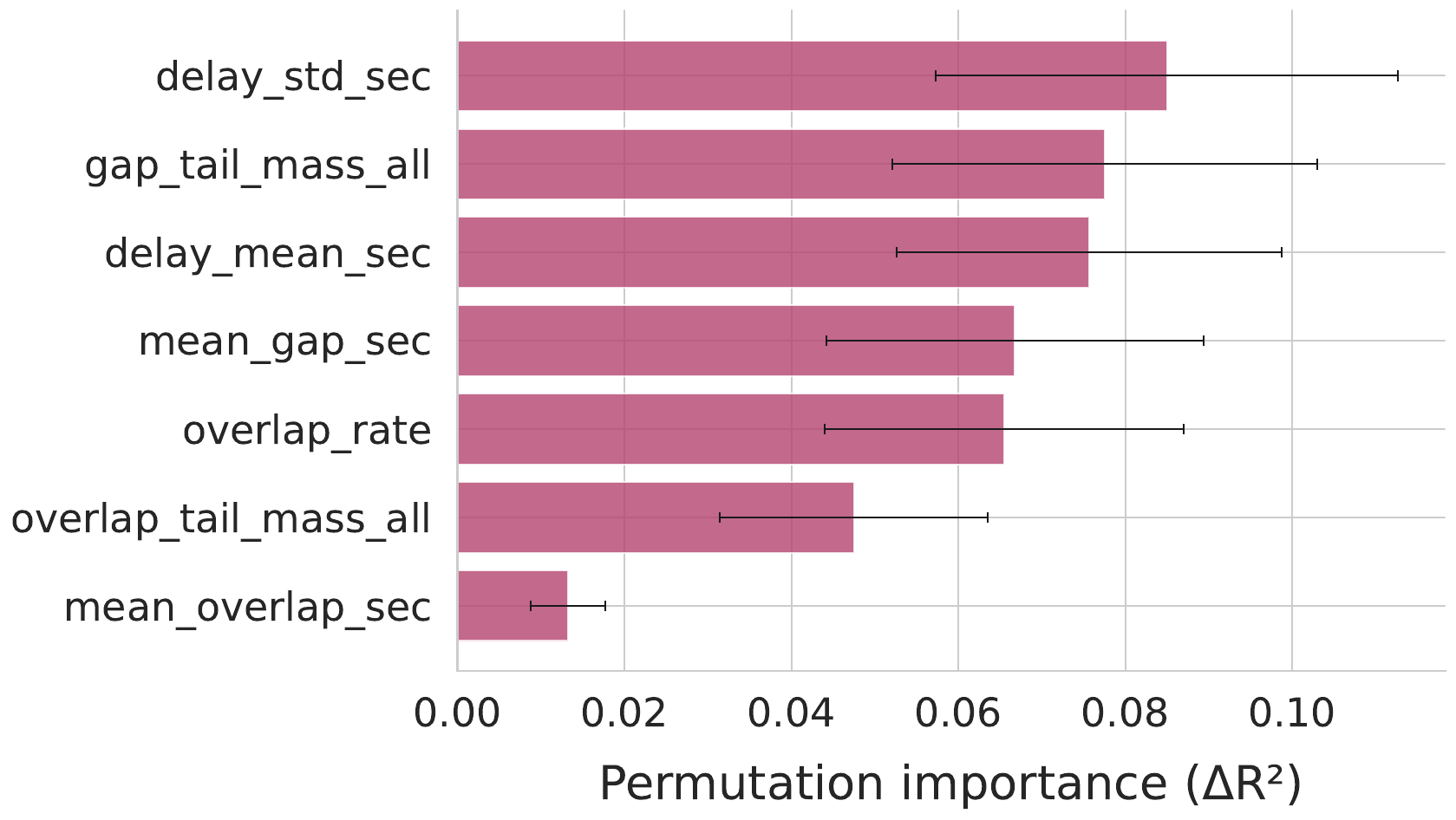}
    \end{minipage}
    \hfill
    \begin{minipage}{0.4\textwidth}
        \centering
        \includegraphics[width=\linewidth]{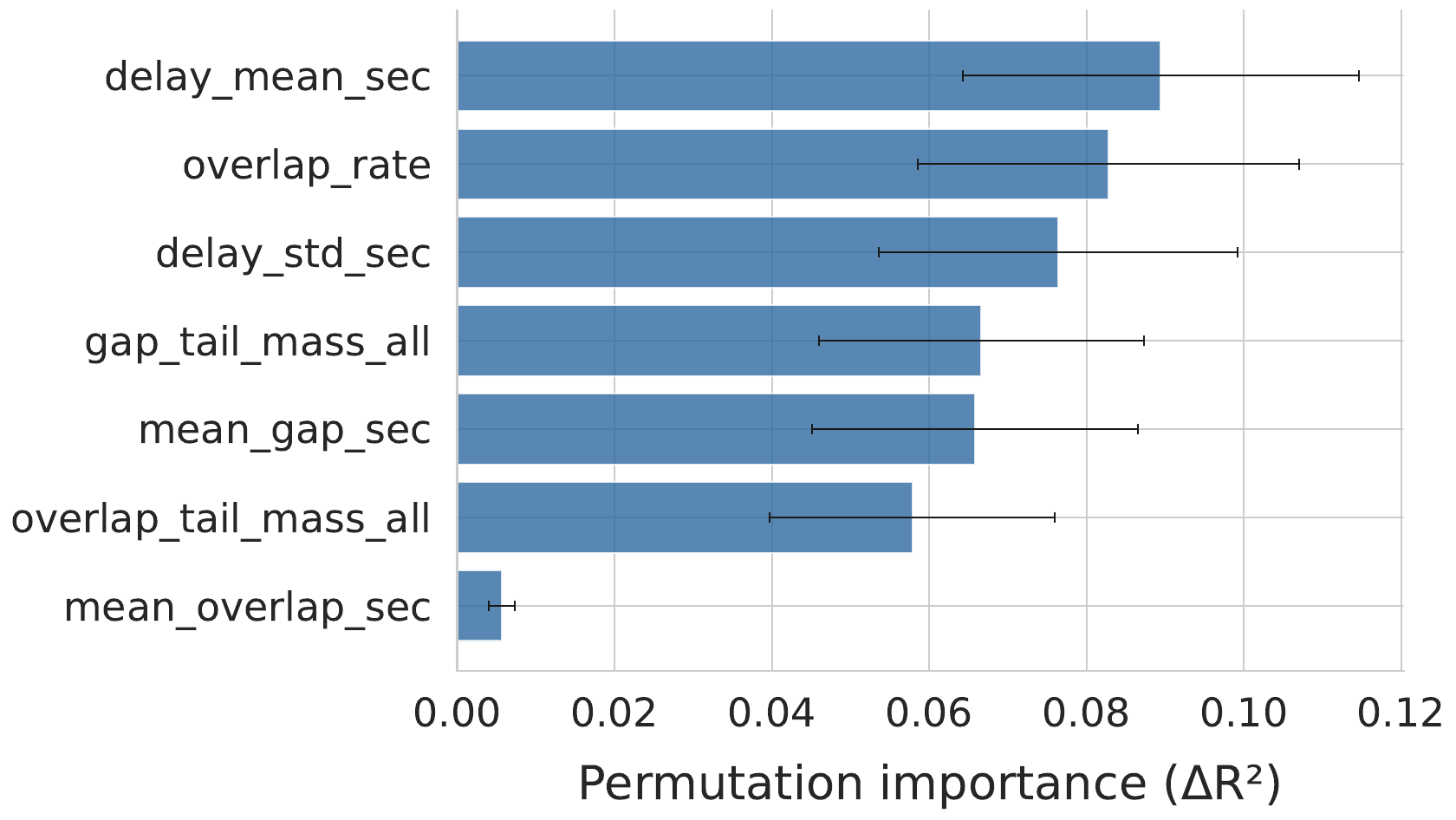}
    \end{minipage}
    \caption{Permutation importance of the timing variables on the evaluation set, shown for cpCER (left) and cpWER (right).}
    \label{fig:timing_relevance_cpwer}
\end{figure*}

\subsection{Timing Statistics Associated With ASR Error (\textit{RQ1})}
\label{sec:results_timing_correlations}

For RQ1, we ask which simple intrinsic timing statistics best explain ASR behavior. As a first step, we use Spearman rank correlation \cite{Spearman1904} because the relationship between timing and recognition error need not be linear. For example, increasing overlap could help up to a point and then saturate; rank correlation still detects the monotonic part of such a trend. Because several timing features are tested at once, we report Benjamini--Hochberg (BH)-adjusted $q$-values \cite{BenjaminiHochberg1995}. A BH-significant result means that the association remains credible after controlling the expected false-discovery rate across the tested timing features. We also checked phase-controlled partial correlations, which remove the LHS/BO phase effect before computing the association; these were used to verify that the main trends are not merely consequences of the optimizer sampling a different region.

Table~\ref{tab:timing_core_eval} reports the results obtained on the evaluation set. The dominant pattern is an overlap--gap trade-off. Configurations with more overlap exposure tend to have lower cpWER, while configurations with longer or more variable gaps tend to have higher cpWER. In particular, overlap rate and overlap tail mass are negatively associated with cpWER, whereas gap tail mass, mean gap, delay mean, and delay variability are positively associated with cpWER. All six associations are BH-significant for cpWER and remain strong after controlling for sampling phase. The cpCER trends point in the same directions, but they are weaker and do not reach BH significance on the evaluation set.

\begin{table}[t]
\centering
\caption{Core evaluation-set timing associations. $\rho$ is Spearman rank correlation with ASR error; $q$ is the BH-adjusted significance value for cpWER.}
\label{tab:timing_core_eval}
\resizebox{0.45\textwidth}{!}{%
\begin{tabular}{lccc}
\toprule
\textbf{Timing feature} & $\rho$ \textbf{(cpWER)} & $q$ \textbf{(cpWER)} & $\rho$ \textbf{(cpCER)} \\
\midrule
Overlap rate & -0.645 & 0.002 & -0.409 \\
Overlap tail mass & -0.622 & 0.003 & -0.399  \\
Mean gap & 0.647 & 0.002 & 0.390  \\
Gap tail mass & 0.656 & 0.002 & 0.409  \\
Delay mean & 0.642 & 0.002 & 0.395 \\
Delay standard deviation & 0.564 & 0.008 & 0.365 \\
\bottomrule
\end{tabular}}

\end{table}

The relevant interpretation is tied to the training distribution: within the explored range, simulations that expose the ASR model to more overlapped and less gap-dominated interaction patterns tend to generalize better on BEA-Dialogue. Mean overlap duration itself is not strongly associated with error, which suggests that the frequency and tail mass of overlap matter more than the average length of overlapping segments. Conversely, the gap-related features move together and consistently predict worse cpWER, indicating that long or highly variable pauses may reduce the usefulness of simulated conversations for this task.

\subsection{Multivariate Predictability and Feature Importance (\textit{RQ1}, \textit{RQ3})}
\label{sec:results_predictive_models}
To quantify how well the intrinsic timing characteristics and the underlying $\theta$ parameters explain downstream ASR performance, we trained three classes of regression models---ridge regression, lasso regression, and random forests---using repeated cross-validation. Models were evaluated separately using (i) intrinsic timing features, (ii) the $\theta$ parameters, and (iii) their combination, with cpWER and cpCER as prediction targets. Because these models are robust to redundant parameterizations, the interpretation of the $\theta$ dimensions extends directly to the corresponding $\phi$ dimensions. Table~\ref{tab:best_predictive_models} reports the best-performing model for each feature group and target according to the cross-validated $R^2$. This is intentionally stricter than in-sample fitting, because with only 25 configurations a flexible model can otherwise appear accurate by overfitting.

\begin{table}[t]
\centering
\caption{Best cross-validated predictive models for each feature group and target metric.}
\label{tab:best_predictive_models}
\resizebox{0.45\textwidth}{!}{%
\begin{tabular}{llccc}
\toprule
\textbf{Feature group} & \textbf{Target} & \textbf{Best model} & $\boldsymbol{R^2_{\mathrm{CV}}}$ & \textbf{RMSE$_{\mathrm{CV}}$} \\
\midrule
\multirow{2}{*}{Timing}
    & cpWER & Random forest & 0.529 & 0.194 \\
    & cpCER & Random forest & 0.247 & 0.092 \\
\midrule
\multirow{2}{*}{$\theta$}
    & cpWER & Ridge & 0.280 & 0.240 \\
    & cpCER & Lasso & 0.064 & 0.103 \\
\midrule
\multirow{2}{*}{Joint}
    & cpWER & Lasso & 0.588 & 0.181 \\
    & cpCER & Random forest & 0.222 & 0.094 \\
\bottomrule
\end{tabular}}
\end{table}

The results consistently demonstrate that intrinsic timing features provide substantially stronger predictive power than the raw $\theta$ coordinates. For cpWER, the timing-only feature set achieves a cross-validated $R^2$ of 0.529 with a random forest model, nearly doubling the predictive power of the best $\theta$-only model ($R^2=0.280$). A similar trend is observed for cpCER, where timing features reach $R^2=0.247$, whereas the $\theta$ representation explains only a negligible fraction of the variance ($R^2=0.064$). Combining the two feature sets yields a modest improvement for cpWER, increasing the best cross-validated performance to $R^2=0.588$ with a lasso model. In contrast, no comparable benefit is observed for cpCER, where the joint feature set achieves $R^2=0.222$, slightly below the timing-only model. The corresponding cross-validated RMSE values mirror these trends, indicating that intrinsic timing statistics capture most of the predictive information available for downstream ASR performance.

These findings suggest that the intrinsic statistics of the generated timing distributions capture most of the information relevant to downstream ASR performance. For cpWER, the gain obtained by the joint model indicates that the $\theta$ parameters still contain complementary information that is not fully reflected by the engineered timing descriptors. This remaining signal may arise from higher-order characteristics of the tilted distributions or interactions between timing properties that are not explicitly represented by the intrinsic feature set.

Permutation importance also helps interpret the fitted models \cite{Breiman2001}. Unlike correlation, which measures how one feature moves with the target in isolation, permutation importance asks how much a trained model suffers when a feature is randomly shuffled while all other features remain fixed. A feature can therefore have high correlation but low permutation importance if it is redundant with related variables. Conversely, a feature can have modest correlation but high importance if it contributes unique information in combination with other features.

Fig.~\ref{fig:timing_relevance_cpwer} shows the permutation importance of the timing variables. The relative ordering differs slightly between the two target metrics; however, mean overlap is consistently the least informative variable by a considerable margin. In both cases, the mean and standard deviation of the full delay distribution rank among the three most important predictors. This is expected, as the overall delay distribution implicitly subsumes the remaining timing descriptors, with its mean and dispersion capturing central aspects of the timing structure. Notably, the two lowest-ranked variables for both metrics are overlap-related. This indicates that performance is not determined by overlap alone; pause characteristics also play an important role.

Table~\ref{tab:theta_phi_importance} compares permutation importance in the raw $\theta$ coordinates and in the identifiable $\phi$ coordinates. In the overcomplete design space, $\theta_2$ is the dominant coordinate for both metrics, with $\theta_3$ contributing a secondary cpWER signal and $\theta_1$/$\theta_4$ carrying little unique importance. Because $\phi_2=\theta_2+\theta_3$ combines the two strongest raw coordinates, one might expect it to become clearly more predictive after compression. The result is more nuanced: $\phi_2$ increases only slightly over $\theta_2$ for cpWER ($0.47$ vs. $0.44$), whereas for cpCER it is noticeably weaker than $\theta_2$ ($0.37$ vs. $0.46$). This is not necessarily contradictory, because $\phi_2$ and $\phi_3=\theta_2+\theta_4$ are coupled through their shared $\theta_2$ component, so compression improves identifiability but can also mix signals that were separable in the overcomplete coordinates. The low importance of $\phi_3$ indicates that this shared component is not sufficient by itself; the gap-side combination represented by $\phi_2$ remains the main identifiable direction. At the same time, $\phi_1=\theta_1$ becomes more visible than $\theta_1$ in the raw model, suggesting that overlap-mass effects are weak but less completely masked once the redundant coordinates are removed. Overall, the comparison supports using both views: $\phi$ gives the cleaner distributional interpretation, while the overcomplete $\theta$ space can preserve useful separations for prediction and optimization.

\begin{table}[t]
\centering
\caption{Mean permutation importance values for the $\theta$ and $\phi$ dimensions. Standard deviations are shown after $\pm$.}
\label{tab:theta_phi_importance}
\resizebox{0.32\textwidth}{!}{%
\begin{tabular}{lcc}
\toprule
\textbf{Dimension} & \textbf{cpWER} & \textbf{cpCER} \\
\midrule
$\theta_1$ & $0.05 \pm 0.02$ & $0.05 \pm 0.01$ \\
$\theta_2$ & $\textbf{0.44} \pm \textbf{0.12}$ & $\textbf{0.46} \pm \textbf{0.14}$ \\
$\theta_3$ & $0.23 \pm 0.08$ & $0.17 \pm 0.06$ \\
$\theta_4$ & $0.05 \pm 0.02$ & $0.06 \pm 0.02$ \\
\midrule
$\phi_1$   & $0.11 \pm 0.05$ & $0.09 \pm 0.03$ \\
$\phi_2$   & $\textbf{0.47} \pm \textbf{0.12}$ & $\textbf{0.37} \pm \textbf{0.10}$ \\
$\phi_3$   & $0.13 \pm 0.06$ & $0.12 \pm 0.05$ \\
\bottomrule
\end{tabular}}
\end{table}

\begin{figure*}[t]
    \centering
    \includegraphics[width=0.8\linewidth]{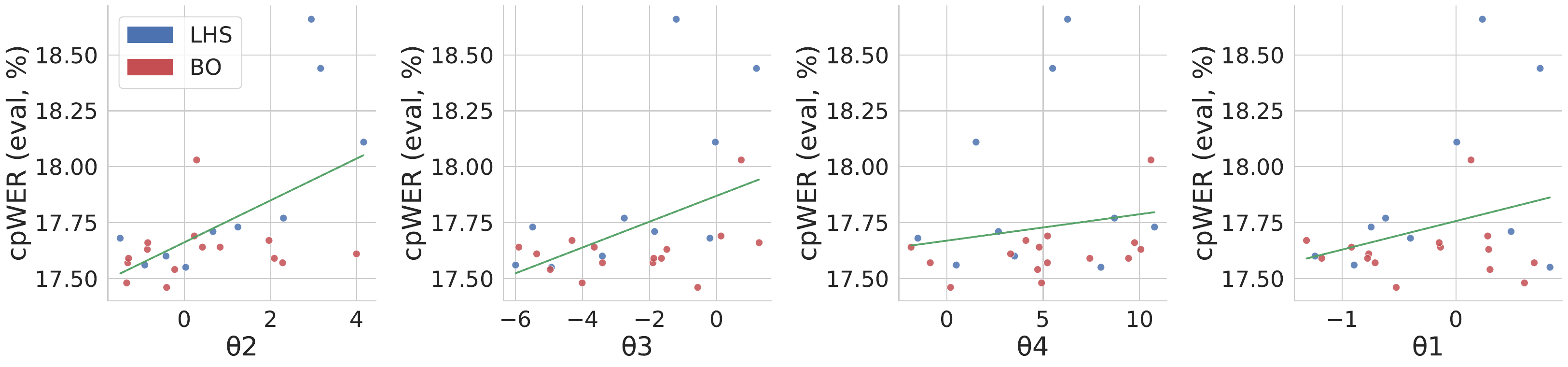}
    \caption{Pairwise $\theta$-space scatter plots with cpWER trends.}
    \label{fig:theta_space_scatter}
\end{figure*}

\subsection{What the Tilt Coordinates Reveal (\textit{RQ3})}
\label{sec:results_theta}

The tilt parameters are important because they are the variables directly manipulated by the simulator. However, as discussed in Section~III, the four-coordinate vector $\theta$ is an overcomplete design representation: the induced density is identifiable through the compressed coordinates $\phi=(\phi_1,\phi_2,\phi_3)$. Table~\ref{tab:theta_phi_assoc_eval} therefore reports both views. The raw $\theta$ coordinates show how the optimizer's intervention variables relate to ASR performance, whereas the $\phi$ coordinates show the same effects after removing the algebraic redundancy of the tilt family.

In the raw design space, $\theta_2$ has the clearest relation to evaluation-set ASR performance: higher $\theta_2$ is associated with higher cpWER ($\rho=0.545$, BH-adjusted $q=0.039$). The next strongest cpWER trend is $\theta_3$ ($\rho=0.459$, $q=0.056$), which is close to but does not pass the BH threshold. The remaining raw coordinates are weaker, and the cpCER associations follow the same broad ordering but with smaller statistical support. This agrees with the permutation importance results in Table~\ref{tab:theta_phi_importance}, where $\theta_2$ and $\theta_3$ are the most informative coordinates and $\theta_1$ contributes little on its own.

The compressed representation sharpens this interpretation. Since $\phi_2=\theta_2+\theta_3$ controls the effective positive-gap slope, its strong association with cpWER ($\rho=0.635$, BH-adjusted $q=0.004$; phase-controlled $\rho_p=0.668$) indicates that the gap side of the tilted distribution is the most informative identifiable direction. The overlap-side slope $\phi_3=\theta_2+\theta_4$ shows a weaker but still consistent cpWER trend ($\rho=0.468$, $q=0.055$), while $\phi_1=\theta_1$, which controls overlap mass directly, is not strongly associated with either metric. Thus, the $\phi$ analysis parallels the raw-coordinate analysis but makes the distributional interpretation clearer: ASR error is most sensitive to how the tilt reshapes the gap side of the timing distribution, with overlap-side effects appearing secondary and overlap-mass shifts alone providing limited explanatory power.

This pattern is consistent with the intrinsic timing analysis. The ASR model does not observe $\theta$ or $\phi$ directly; it observes generated conversations. The parameter-space results therefore support, rather than replace, the timing-statistic results: raw $\theta$ coordinates are useful for diagnosing the optimizer's intervention space, $\phi$ coordinates are useful for identifiable distributional interpretation, and the generated overlap, gap, and variability statistics remain the most direct explanation of downstream ASR behavior.

\begin{table}[t]
\centering
\caption{Associations between the $\theta$ and $\phi$ parameters and evaluation-set ASR performance. $\rho$ denotes Spearman's rank correlation, $\rho_p$ the phase-controlled partial Spearman correlation, and $q$ the Benjamini--Hochberg adjusted significance value.}
\label{tab:theta_phi_assoc_eval}
\resizebox{0.49\textwidth}{!}{%
\begin{tabular}{lccccc}
\toprule
\textbf{Parameter} &
$\rho$ \textbf{(cpWER)} &
$\rho_p$ &
$q$ &
$\rho$ \textbf{(cpCER)} &
$\rho_p$ \\
\midrule
$\theta_1$ &  0.057 &  0.014 & 0.836 & -0.056 & -0.186 \\
$\theta_2$ & \textbf{0.545} & \textbf{0.517} & \textbf{0.039} & \textbf{0.462} & \textbf{0.463} \\
$\theta_3$ &  0.459 &  0.509 & 0.056 &  0.215 &  0.308 \\
$\theta_4$ &  0.256 &  0.305 & 0.434 &  0.044 &  0.109 \\
\midrule
$\phi_1$   &  0.057 &  0.014 & 0.790 & -0.056 & -0.186 \\
$\phi_2$   & \textbf{0.635} & \textbf{0.668} & \textbf{0.004} &  0.395 &  0.486 \\
$\phi_3$   &  0.468 &  0.497 & 0.055 &  0.241 &  0.299 \\
\bottomrule
\end{tabular}}
\end{table}

Fig.~\ref{fig:theta_space_scatter} visualizes cpWER as a function of each individual $\theta$ parameter. To aid interpretation, a linear trend and a polynomial regression curve are fitted in this and all following figures. For the latter, the order of the polynomial regression is chosen based on the Akaike information criterion (AIC). When the selected polynomial order is one, only a single trend line is shown. For $\theta_2$ and, to a lesser extent, $\theta_1$, the largest cpWER values occur near the upper end of the sampled range, suggesting that excessively large values along these dimensions may be associated with degraded ASR performance. In contrast, the relationship for $\theta_1$ is less straightforward: while several of the highest-error experiments occur at large $\theta_1$ values, the same region also contains some of the best-performing configurations. This indicates that $\theta_1$ alone is not a strong determinant of downstream performance, and that its effect depends on the values of the remaining $\theta$ parameters. This is consistent with Table~\ref{tab:theta_phi_importance}, which shows that $\theta_1$ is comparatively less predictive than $\theta_2$ or $\theta_3$.

\subsection{Corpus Similarity and Geometry of the Timing Space (\textit{RQ2}, \textit{RQ4})}
\label{sec:results_geometry}

\begin{figure*}[t]
    \centering
    \includegraphics[width=0.9\linewidth]{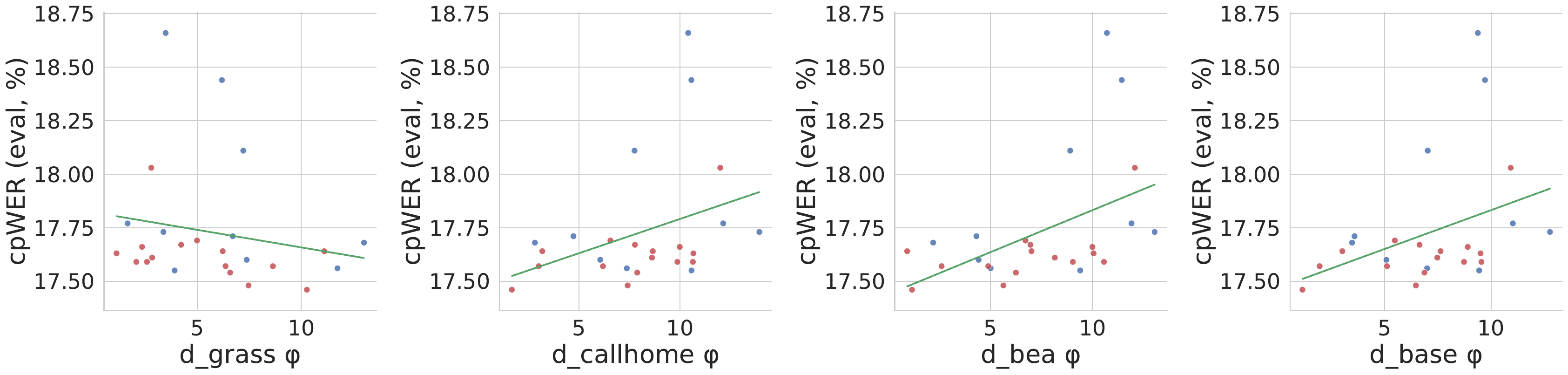}
    \caption{ASR error as a function of distance to reference timing embeddings in $\phi$ space. The plot tests whether low-error settings are corpus-like or extrapolated from the reference corpora.}
    \label{fig:theta_distance_scatter}
\end{figure*}

We next examine whether high-performing configurations are simply those that lie closest to real-corpus timing profiles. Fig.~\ref{fig:theta_distance_scatter} compares ASR error against distances to the reference corpus embeddings in $\phi$ space. We use $\phi$ rather than $\theta$ for this analysis because Euclidean distances in the redundant $\theta$ parameterization are less directly interpretable: the same induced timing distribution may admit multiple $\theta$ embeddings. This diagnostic therefore separates corpus resemblance from task relevance. Although some linear trends are visible, these are driven largely by configurations at the largest distances; once these extreme cases are excluded, the relationship between corpus distance and ASR error becomes substantially weaker and less consistent. Together with the preceding analyses, this suggests that ASR behavior is more directly explained by the induced overlap--gap statistics than by proximity to a particular corpus reference. Corpus similarity is therefore useful for characterizing where a configuration lies in the search space, but it is not, by itself, sufficient to explain performance.

\begin{figure*}[t]
    \centering
    \includegraphics[width=0.9\linewidth]{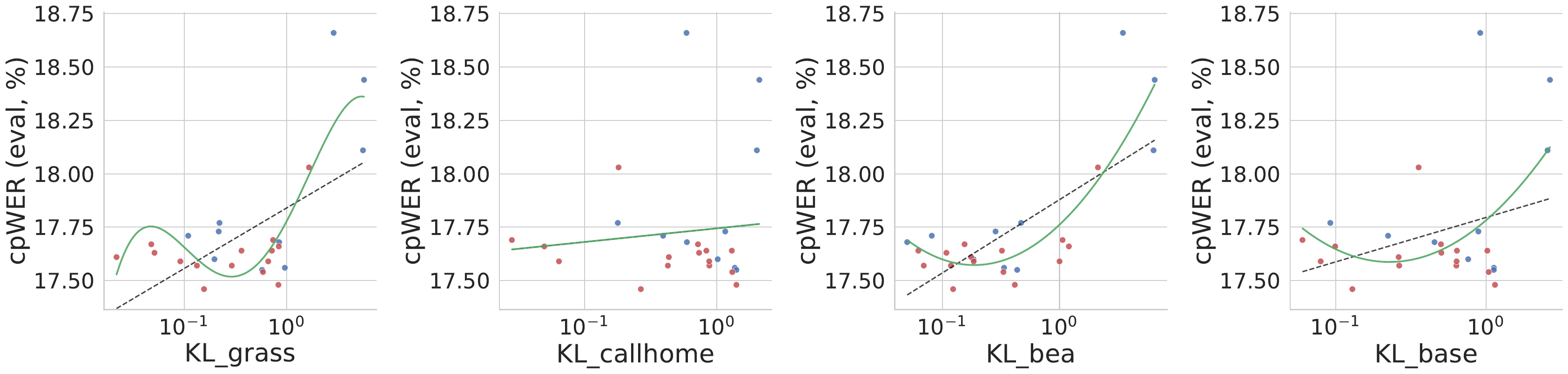}
    \caption{ASR error as a function of KL divergence to reference timing distributions.}
    \label{fig:theta_kl_scatter}
\end{figure*}

Fig.~\ref{fig:theta_kl_scatter} presents the corresponding analysis using Kullback--Leibler (KL) divergence instead of Euclidean distance in $\phi$ space. Because the KL values contain extreme outliers, the divergence values are shown on a logarithmic scale. The resulting point configurations and fitted trends differ markedly from those obtained with Euclidean distances, suggesting that coordinate-space proximity does not necessarily preserve distributional similarity. In contrast, the KL-based representation reveals a clearer relationship between distributional deviation and downstream performance: experiments that diverge strongly from the real-corpus distributions tend to yield higher ASR error. This indicates that excessively aggressive manipulation of the timing distribution is unlikely to benefit training, and that remaining within a distributionally plausible region is important for effective augmentation.

Finally, principal component analysis (PCA) summarizes how much redundancy exists in the sampled spaces \cite{Jolliffe2002PCA}. For the four $\theta$ variables, PC1 explains 33.2\% of the variance and the first two PCs explain 61.0\%. This means that the sampled $\theta$ space is not effectively one-dimensional; multiple tilt degrees of freedom are active. In contrast, the seven timing features are much more coordinated: PC1 explains 83.3\% of the variance and the first two PCs explain 96.9\%. This supports a useful interpretation of the whole analysis: the simulator is controlled by several $\theta$ directions, but many of their observable consequences collapse onto a dominant overlap--gap axis. For the $\phi$ variables, PC1 explains 48.3\% of the variance, and PC1 and PC2 together explain 80.4\%.

\begin{figure}[h]
    \centering
    \includegraphics[width=0.95\linewidth]{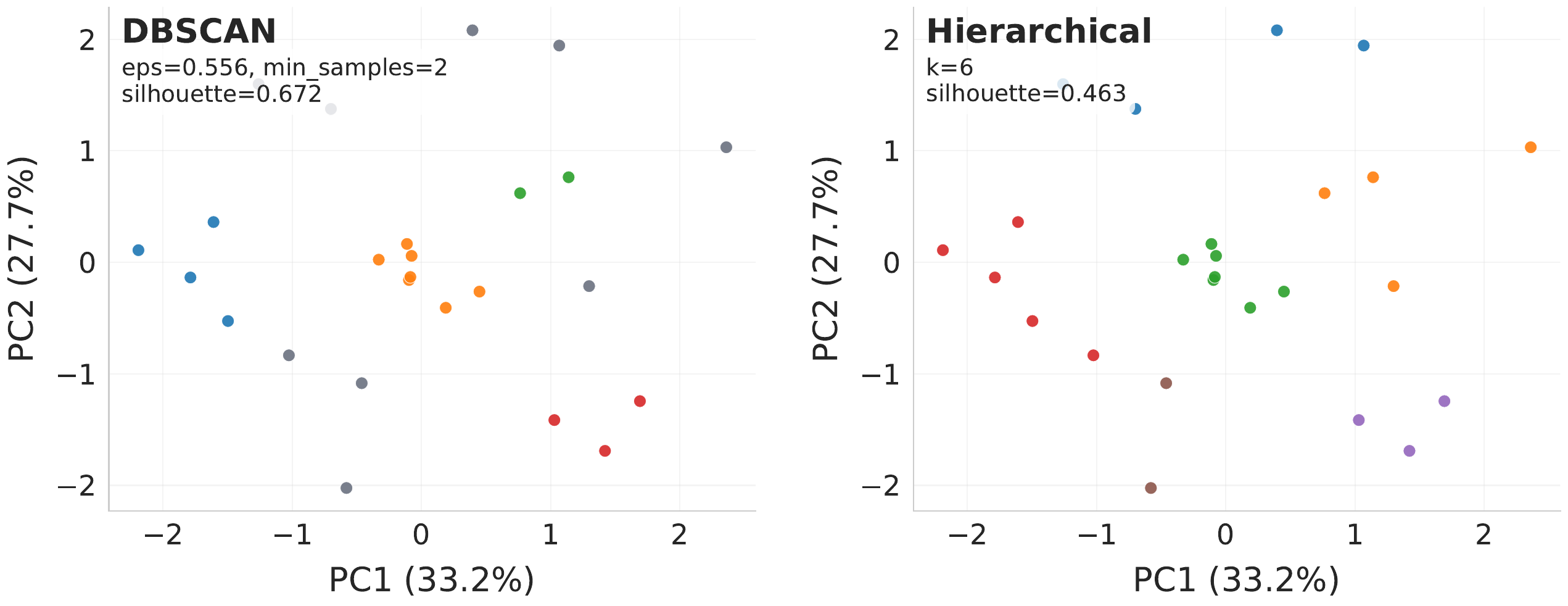}
    \caption{PCA and clustering view of the sampled $\theta$ space. The first two PCs capture 61.0\% of $\theta$ variance, indicating partial but not complete redundancy among tilt dimensions.}
    \label{fig:pca_clustering}
\end{figure}

To explore whether different regions of the sampled $\theta$ space exhibit distinct performance characteristics, we clustered the two-dimensional PCA representation shown in Fig.~\ref{fig:pca_clustering} using both DBSCAN and agglomerative hierarchical clustering. Table~\ref{tab:pca_clusters} summarizes the clusters with the lowest and highest mean ASR error identified by each method.

Although the cluster assignments differ between the two algorithms, both partitions identify groups with noticeably different average cpWER and cpCER values. The clusters with the lowest mean error achieve cpWER values of approximately 17.58\%, whereas the highest-error clusters exceed 18.1\%. The higher-error clusters also exhibit larger within-cluster standard deviations, although some of these clusters contain only a small number of experiments. Consequently, these observations should be interpreted as exploratory rather than conclusive. Nevertheless, the clustering analysis suggests that neighboring regions in the PCA representation can exhibit similar downstream performance, indicating that the sampled $\theta$ space contains localized regions with characteristic timing behavior.

\begin{table}[t]
\centering
\caption{Best- and worst-performing clusters in the PCA representation of the $\theta$ space. Values denote mean $\pm$ sample standard deviation.}
\label{tab:pca_clusters}
\resizebox{0.45\textwidth}{!}{%
\begin{tabular}{llccc}
\toprule
\textbf{Method} & \textbf{Cluster} & \textbf{$n$} & \textbf{cpWER} & \textbf{cpCER} \\
\midrule
DBSCAN & 1 (best) & 7 & $17.581 \pm 0.110$ & $8.169 \pm 0.066$ \\
DBSCAN & 2 (worst) & 2 & $18.185 \pm 0.672$ & $8.395 \pm 0.233$ \\
\midrule
Hierarchical & 5 (best) & 2 & $17.580 \pm 0.014$ & $8.150 \pm 0.028$ \\
Hierarchical & 1 (worst) & 4 & $18.125 \pm 0.499$ & $8.355 \pm 0.185$ \\
\bottomrule
\end{tabular}}
\end{table}

\subsection{Comparison of Representative ASR Metrics (\textit{RQ2}, \textit{RQ5})}
\label{sec:results_asr_metric_comparison}

The correlation and feature-importance analyses above describe broad trends across the explored timing space. To put these trends in perspective, Table~\ref{tab:representative_asr_metrics} compares the corpus-derived reference configurations, using results reported in \cite{C-SASC}, against selected sampled configurations. This table is not intended as a significance test; rather, it shows the scale of the observed differences and whether the best sampled configurations clearly separate from corpus-like timing baselines.

\begin{table*}[t]
\centering
\caption{Representative ASR results for corpus-derived references and selected sampled configurations. Lower cpWER/cpCER values are better.}
\label{tab:representative_asr_metrics}
\begin{tabular}{llcccc}
\toprule
\textbf{Configuration} & \textbf{Description} & \textbf{Dev cpWER} & \textbf{Dev cpCER} & \textbf{Eval cpWER} & \textbf{Eval cpCER} \\
\midrule
BEA & Corpus-derived reference \cite{C-SASC} & 16.25 & 6.91 & 17.64 & 8.13 \\
CallHome & Corpus-derived reference \cite{C-SASC} & 16.23 & 6.82 & 17.76 & 8.16 \\
GRASS & Corpus-derived reference \cite{C-SASC} & 16.21 & 6.83 & 17.63 & 8.11 \\
\midrule
Tilt Base & Base density (weighted mixture) & 16.26 & 6.89 & 17.52 & 8.08 \\
\midrule
LHS-9 & Best LHS configuration by development cpWER & \textbf{16.05} & 6.82 & 17.63 & 8.15 \\
LHS-2 & Best LHS configuration by evaluation cpWER & 16.13 & 6.84 & \textbf{17.44} & \textbf{8.07} \\
BO-20 & Best BO configuration by development cpWER/cpCER & \textbf{16.05} & \textbf{6.74} & 17.54 & 8.11 \\
BO-12 & Best BO configuration by evaluation cpWER & 16.41 & 6.85 & 17.46 & 8.15 \\
BO-13 & Best BO configuration by evaluation cpCER & 16.28 & 6.87 & 17.48 & 8.09 \\
\bottomrule
\end{tabular}
\end{table*}

Several conclusions follow from this comparison. First, the base-distribution experiment defined in \eqref{eq:base_mixture} performs better than the corpus-derived references on the evaluation set, but worse on the development set. Second, the sampled configurations improve on the corpus-derived references in some cases, but the absolute differences are small. For example, the best evaluation-set cpWER among the listed configurations is 17.44, compared with 17.63--17.76 for the corpus references. The best listed evaluation-set cpCER is 8.07, compared with 8.11--8.16 for the corpus references. These are useful differences, but they are not large enough to make a best-configuration narrative convincing by itself.

Third, the configuration that performs best on the development set is not necessarily the best on the evaluation set. BO-20 has the best development cpCER and tied-best development cpWER among the selected sampled configurations, but it is not the best configuration on the evaluation set. This agrees with the development--evaluation analysis in Fig.~\ref{fig:dev_eval_consistency}: development metrics capture broad tendencies, but the exact ordering of configurations is unstable. As a result, the more reliable contribution of the experiment is the identification of timing properties that repeatedly covary with error, rather than the selection of one nominally optimal setting.

\begin{figure}[t]
    \centering
    \includegraphics[width=0.75\linewidth]{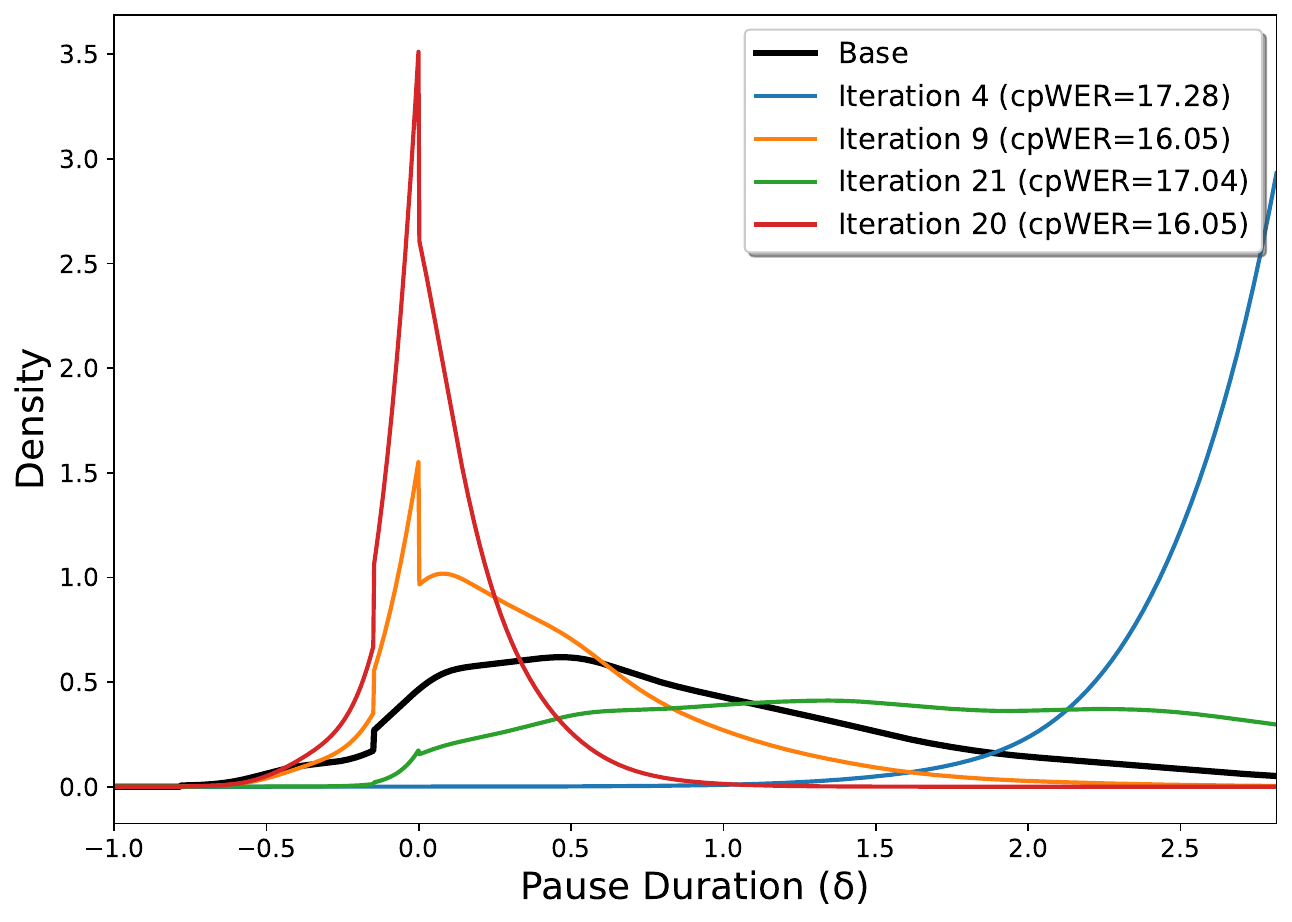}
    \caption{Representative tilted timing distributions for selected configurations illustratinh how low- and high-error settings differ in overlap/gap balance.}
    \label{fig:representative_tilted_distributions}
\end{figure}

Finally, the representative timing distributions in Fig.~\ref{fig:representative_tilted_distributions} show that low-error settings do not require implausible timing shapes. The selected configurations remain within the family of corpus-like distributions but shift the balance between overlap and gap behavior. This is consistent with the main statistical finding: useful configurations are not arbitrary distortions of conversational timing, but controlled redistributions that increase overlap exposure and reduce long-gap dominance within the explored range.

\subsection{Summary of Main Findings}
\label{sec:results_summary}

The results give coherent answers to the research questions (\textit{RQ1}--\textit{RQ5}). BO provides modest evaluation-set improvement, but the main value of the framework is analytical: it creates controlled timing configurations that reveal which properties matter. The strongest evidence points to the induced timing statistics rather than to the raw parameter vector. Configurations with more overlap exposure and fewer long gaps tend to reduce cpWER, while gap-heavy and highly variable timing distributions tend to increase it. cpCER follows the same direction, but with weaker statistical support. The $\theta$-space analysis shows that several tilt dimensions are active, yet their observable timing effects largely align along an overlap--gap axis. The direct ASR comparison further shows that sampled configurations can improve over corpus-derived references, but only by modest margins; this reinforces the interpretation that the contribution is not a single winning setting, but a better understanding of why some timing regimes are more useful than others. For simulation design, this means that monitoring the generated timing distribution is essential: choosing $\theta$ is useful only insofar as it produces the overlap, gap, and variability profile needed by the ASR task.

\section{Discussion}

The central outcome of this study is that the usefulness of simulated conversations for ASR is not fully explained by corpus realism alone (\textit{RQ2}). Corpus-derived configurations provide sensible and stable reference points, but the best sampled settings in Table~\ref{tab:representative_asr_metrics} slightly outperform them by shifting the generated timing distribution within the same broad family. Thus, realism is a useful starting point, but the training distribution can also be shaped toward timing conditions that are especially informative for the target ASR task.

The most consistent empirical pattern is the overlap--gap trade-off (\textit{RQ1}). Across correlation, predictive modeling, and permutation-importance analyses, configurations with more overlap exposure and less dominance by long gaps tend to reduce cpWER. This does not imply that arbitrary or excessive overlap should be added. Rather, within the explored range, overlap exposes the recognizer to relevant multi-speaker regions, whereas long and variable gaps may allocate too much simulation budget to acoustically easier or less informative regions. The weaker cpCER trends suggest that this effect is more visible at the word level, possibly because word-level errors are more sensitive to segmentation, ordering, and speaker-interaction ambiguity.

A second implication is that the raw tilt vector is less interpretable than the timing distribution it induces (\textit{RQ3}--\textit{RQ4}). The optimizer acts in the four-dimensional $\theta$ space, but the ASR model is trained on generated conversations, not on $\theta$ itself. This explains why intrinsic timing features predict cpWER better than the raw coordinates, and why adding $\theta$ provides only a limited cross-validated gain. In practice, simulator parameters should therefore be monitored through observable consequences such as overlap rate, gap tail mass, delay mean, and delay variability, which are more actionable than coordinate values alone.

The compressed $\phi$ space refines this point by separating identifiable directions in the tilted density from convenient optimizer coordinates. The strongest and most stable $\phi$ result is the role of $\phi_2=\theta_2+\theta_3$, which captures the effective positive-gap-side tilt and aligns with the broader finding that gap behavior is central to cpWER. However, the compression does not uniformly improve on the raw coordinates: $\phi_2$ only slightly improves on $\theta_2$ for cpWER and is weaker for cpCER, while $\phi_3=\theta_2+\theta_4$ remains comparatively unimportant despite sharing the dominant $\theta_2$ component. Thus, identifiable directions clarify the distributional interpretation but can also merge signals that the overcomplete $\theta$ representation keeps separable for optimization. The increased apparent importance of $\phi_1=\theta_1$ further suggests that overlap-mass effects are present but secondary. Future work should test whether simulators should search directly in the more interpretable $\phi$ space or retain an overcomplete parameterization, and whether the weak $\phi_3$ and $\phi_1$ effects are genuinely small or simply difficult to estimate with the available budget.

The Bayesian optimization phase is useful, but its role should be interpreted carefully (\textit{RQ5}). It produces a small evaluation-set improvement and moves the search toward more favorable regions, yet the development set does not show the same median improvement and rank agreement across splits is only moderate. These observations argue against presenting the method as finding a single optimal timing setting. Its stronger contribution is methodological: the LHS--BO loop creates controlled timing interventions that reveal which properties of the generated data matter, so optimization serves both as a search tool and as an experimental-design mechanism.

Several limitations remain. The number of evaluated configurations is small because each point requires full simulation and ASR training, so the observed patterns should be confirmed with larger budgets and repeated runs. The downstream evaluation is centered on BEA-Dialogue, and the preferred overlap--gap balance may differ across languages, recording conditions, segmentation conventions, or ASR architectures. In addition, this work varies only the mean timing distribution while keeping the residual timing model fixed, leaving speaker-specific dynamics, lexical content, acoustic environment, and correlations between timing and speaker behavior untested. Finally, the aggregate timing summaries used here should be connected more directly to error types in overlapped, near-boundary, and long-silence regions.

These limitations motivate applying the framework to additional corpora and recognizers and extending the parameterization to speaker- or context-conditioned timing controls. More broadly, the results suggest that synthetic conversation generation should be evaluated not only by distributional similarity to real corpora, but also by task-relevant diagnostics of how generated overlap, gap, and variability profiles affect downstream learning.

\section{Conclusion}

This paper studied what makes simulated conversational timing useful for ASR training. We proposed an analysis framework based on exponential tilting of corpus-derived timing densities, explored the resulting parameter space with LHS and multi-objective Bayesian optimization, and evaluated each timing configuration through simulated training data and downstream cpWER/cpCER.

The main finding is that induced timing statistics explain ASR behavior better than either corpus identity or raw simulator coordinates. In the explored region, configurations with more overlap exposure and fewer long or highly variable gaps tend to reduce cpWER, whereas gap-heavy timing profiles tend to increase it. Bayesian optimization yields modest improvements over the initial sampled configurations and over corpus-derived references, but the larger value of the approach is explanatory: it reveals which properties of simulated timing distributions are associated with better recognition performance.

Overall, the results support treating conversational timing as a controllable training variable rather than as a fixed corpus statistic to be copied exactly. Realistic simulation remains important, but task usefulness depends on the particular overlap, gap, and variability profile produced by the simulator. Future work should validate these patterns across more corpora and ASR systems, extend the parameterization to richer speaker- and context-dependent timing behavior, and connect timing interventions to more detailed analyses of recognition errors in conversational speech.

\section*{Acknowledgments}
This work was supported by the Ministry of Culture and Innovation of Hungary from the National Research, Development and Innovation Fund under Project No. 2025-2.1.2-EKÖP-KDP-2025-00005, financed through the EKÖP\_KDP-25-1-BME-21 funding scheme. Additional support was provided by the Hungarian NRDI Fund under projects NKFIH K143075, K135038, and NKFIH-828-2/2021 (MILAB).


\bibliographystyle{IEEEtran}
\bibliography{IEEEabrv,main}

 


\vfill

\end{document}